\documentclass[aps,twocolumn,pra,superscriptaddress,amsmath,showpacs,tightenlines]{revtex4}
\usepackage{epsfig,graphicx,times}
\usepackage{amstext}
\usepackage{amsmath}            
\usepackage{amssymb}            
\usepackage{graphicx}           
\usepackage{latexsym}
\usepackage{bm}
\begin{document}
\title{Coexistence of single- and multi-photon processes due to longitudinal
couplings between superconducting flux qubits and external fields}

\author{Yu-xi Liu}
\affiliation{Institute of Microelectronics, Tsinghua University,
Beijing 100084, China} \affiliation{Tsinghua National Laboratory for
Information Science and Technology, Beijing
100084, China}

\author{Cheng-Xi Yang}
\affiliation{Department of Physics, Tsinghua University, Beijing
100084, China}

\author{Hui-Chen Sun}
\affiliation{Institute of Microelectronics, Tsinghua University,
Beijing 100084, China}

\author{Xiang-Bin Wang}
\affiliation{Department of Physics, Tsinghua University, Beijing
100084, China}
\date{\today}

\begin{abstract}

In contrast to natural atoms, the potential energies  for
superconducting flux qubit (SFQ) circuits can be artificially
controlled. When the inversion symmetry of the potential energy is
broken, we find that the multi-photon processes can coexist in
the multi-level SFQ circuits. Moreover, there are not only
transverse but also longitudinal couplings between the external
magnetic fields and the SFQs when the inversion symmetry of
potential energy is broken. The longitudinal coupling would induce
some new phenomena in the SFQs. Here we will show how the
longitudinal coupling can result in the coexistence of multi-photon
processes in a two-level system formed by a SFQ circuit. We also
show that the SFQs can become transparent to the transverse coupling
fields when the longitudinal coupling fields satisfy the certain
conditions. We further show that the quantum Zeno effect can also be
induced by the longitudinal coupling in the SFQs. Finally we clarify
why the longitudinal coupling can induce coexistence and
disappearance of single- and two-photon processes for a driven SFQ,
which is coupled to a single-mode quantized field.

\end{abstract}

\maketitle \pagenumbering{arabic}

\section{Introduction}
Superconducting quantum circuits possess discrete energy-levels and
behave like natural atoms~\cite{p1,r1,r2,r3,r4,r5} that the
transitions between different energy levels can be induced by the
microwave electromagnetic fields. Thus, many experiments implemented
in natural atoms can also be demonstrated by using the
superconducting quantum circuits, e.g., circuit quantum
electrodynamics (e.g., in Refs.~\cite{wallraff,mooij1,review1}),
superconducting qubit dressed-states (e.g., in
Refs.~\cite{liudressed,large-charge3,Wilson-PRB,wallraff2009,greenberg})
and microwave amplification~\cite{jena}, state control in
superconducting quantum three-level systems (e.g., in
Refs.~\cite{yang1,yang2,orlando2004,orlando2006,nist,goan,falci,falci-1,ian,Abdumalikov}),
lasing without population inversion~\cite{blais,wei},  cooling for
the superconducting qubits (e.g., in
Refs.~\cite{youjqprl,orlandon,cooling}), and sideband excitations
(e.g., in Refs.~\cite{side1,side2,side3}). In contrast to natural atoms, the
potential energy of superconducting flux qubit (SFQ) circuits can be
changed by adjusting externally applied magnetic fields. The
tunability of potential energy makes the SFQ circuits have many features which do not
exist in natural atoms.

In natural atoms, the inversion symmetry of the potential energy is
given by the nature and cannot be changed artificially. Thus each
eigenstate has well-defined parity, and the electric-dipole
transitions induced by the electric field can only link two
eigenstates which have different parities. However, the inversion
symmetry of the potential energy for the SFQ circuits can be
controlled by the external magnetic flux. When the inversion
symmetry is adjusted to be broken, there is no well-defined parity
for each eigenstate of the multi-level SFQ circuits, and the
microwave-induced transitions between any two energy levels are
possible, thus the multi-photon and single-photon processes can
coexist for such multi-level systems. This coexistence of the
multi-photon processes can be easily understood by virtue of an
example using three-level SFQ circuits. That is, the transition
between the ground state and the second excited state can be
realized via two different pathes: (i) from the ground to the second
excited state via a single-photon process; or (ii) from the ground
state to the first excited state via a single-photon, and then from
the first excited state to the second excited state via another
single-photon process~\cite{liu2005}. This means that the single-
and two-photon processes can realize the same goal: the transition
from the ground to the second excited state.

The coexistence~\cite{liu2005} of the single- and two-photon
processes in the three-level SFQ circuits with the cyclic
transition, which is also called as $\Delta$-type transition in
analogue to so-called $\Xi$-, $\Lambda$-, and $V$-type transitions
in atomic physics or quantum optics~\cite{scullybook}, has been
experimentally demonstrated via a delicate superconducting
qubit-resonator circuit~\cite{naturephysics2008,nano}. The
three-level SFQ circuits with $\Delta$-type transitions can be used
to generate single photon~\cite{youjqprb,nec} and cool
superconducting qubits~\cite{youjqprl,orlandon}. The analysis of the
inversion symmetry for the potential energy of the
SFQs~\cite{liu2005,liu2006} also showed that two SFQs cannot simultaneously work at
the optimal point when the frequency matching method is used to
control the coupling between them. Thus either an auxiliary circuit or
a coupler~\cite{side1,plourde,plourde-1,miro1} is necessary to make
that both of the SFQs can be at their optimal
points~\cite{side1,liu2006}. Afterwards, several theoretical
works~\cite{mooij,tsai,miro,franco} followed proposals in
Refs.~\cite{side1,liu2006} and studied how to control the two-flux-qubit
coupling using an additional nonlinear coupler, which resulted in
experimental studies on controllable couplings~\cite{tsaiscience} and
engineered selection rules for tunable couplings~\cite{harrabi}.

In this work, we will first show that the transverse and
longitudinal couplings between the microwave fields and the SFQs can
coexist. Such coexistence results from the broken inversion symmetry of
tunable potential energy of the SFQ circuits. The transverse
coupling between a single-mode microwave field and the SFQ, which
can be reduced to the Jaynes-Cummings model in the rotating wave
approximation, is well studied in the quantum optics and atomic
physics~\cite{scullybook}. However, the model with both the
transverse and longitudinal couplings is less studied. The reason is
that electric-dipole induced longitudinal couplings do not exist in
the natural atoms with well-defined parities. In this paper, we show some
new results, which do not exist in the Jaynes-Cummings model, when both transverse and longitudinal couplings exist.
Moreover, we also study the interaction between the driven SFQs
and the low frequency harmonic
oscillator~\cite{cooling,shnirman,shnirman-1} when the inversion
symmetry of the SFQs is broken, and further show some new phenomena due to
the longitudinal coupling.

Our paper is organized as below. In Sec.~II, we will briefly review
the SFQ circuits. As the complementary and generalization of
Ref.~\cite{liu2005}, we will also clarify some points which were not
studied in our earlier literatures. For instance, how to take phase
transformations so that the interaction between the SFQ circuits and
the external circuit can be described via the product of the loop
current of the SFQ circuits and the external magnetic flux, and how the
multi-photon processes can coexist in multi-level systems when the
inversion symmetry is broken. In Sec.~III, we will present a
Hamiltonian on the transverse and longitudinal couplings between the
magnetic fields and SFQs, and show how the longitudinal coupling can
induce the coexistence of single- and multi-photon processes in a
superconducting quantum two-level system. We will also demonstrate
the longitudinal coupling induced dynamical quantum Zeno effect and
the transparency of the SFQs to the transverse coupling fields. In
Sec.~IV, the transverse and longitudinal couplings between the
driven SFQ and the low frequency harmonic oscillator (e.g., an LC
circuit) is studied. We will explore the nature of the coexistence
and disappearance of the single- and two-photon processes in the
driven SFQs. Finally, we summarize our results in Sec.~V.

\begin{figure}
\includegraphics[bb=20 15 280 210, width=8 cm, clip]{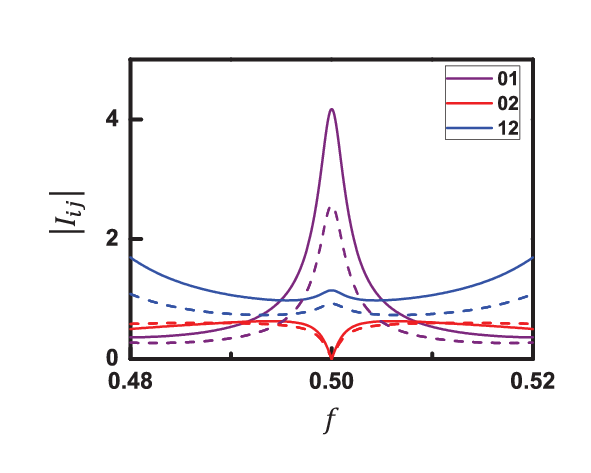}
\caption[]{(Color online) Comparisons of the moduli of the
$f$-dependent transition matrix elements $I_{ij}=\langle i|I|j\rangle$ with
$i\neq j$ between the loop current in Eq.~(\ref{eq:3}) with solid
curves and those in Ref.~\cite{liu2005} with dashed curves for
the three lowest energy levels with $i,\,j=0,\,1,\,2$. Here, the
different transition matrix elements are denoted by different colors
as shown in the figure. We take $\alpha=0.8$ and $E_{\rm J}=40
E_{\rm c}$ in our numerical simulations. Here, $E_{\rm c}=e^2/(2C_{\rm J})$.} \label{fig1}
\end{figure}

\section{Theoretical model and coexistence of multi-photon processes in multi-level SFQ circuits}

In this section, we first study theoretically the interaction model
between the SFQ circuit with three Josephson junctions and the
externally applied time-dependent magnetic flux.
In the following, the abbreviation SFQ usually denotes two-level system (qubit) of
the SFQ circuit if we do not specify it. As the
complementary and generalization of the results in
Ref.~\cite{liu2005}, we will clarify some points which were not
studied in the former literatures (e.g., in
Refs.~\cite{liu2005,orlando}). Then we will summarize the selection
rules and discuss the coexistence of multi-photon processes in the
multi-level systems.

\subsection{Hamiltonian and phase transformations}

Let us consider a superconducting flux qubit (SFQ) circuit, which is
composed of a superconducting loop with three Josephson junctions.
As in Ref.~\cite{liu2005} and Ref.~\cite{orlando}, the two larger
junctions are assumed to have equal Josephson energies $E_{\rm{J1}
}=E_{\rm{J2}}=E_{\rm{J}}$ and capacitances
$C_{\rm{J1}}=C_{\rm{J2}}=C_{\rm{J}}$. While for the third junction,
the Josephson energy and the capacitance are assumed to be $E_{\rm
J3}=\alpha E_{\rm J}$ and $C_{\rm J3}=\alpha C_{\rm J}$, with
$0.5<\alpha <1$. We assume that a static magnetic flux $\Phi_{e}$
and a time-dependent magnetic flux $\Phi(t)$ are applied through the
superconducting loop. In this case, the Hamiltonian can be given by
\begin{equation}\label{eq:1}
H=\frac{P_{p}^{2}}{2M_{p}}+\frac{P_{m}^{2}}{2M_{m}}+U(\varphi
_{p},\varphi _{m})+I\Phi(t),
\end{equation}
with $M_{p}=2C_{\rm J}(\Phi_{0}/2\pi)^2$ and $M_{m}=M_{p}(1+2\alpha)$.
The potential energy $U(\varphi _{p},\varphi _{m})$ of the SFQ
circuit is defined as
\begin{eqnarray}\label{eq:2u}
U(\varphi _{p},\varphi _{m})& =& 2E_{\rm{J}}(1-\cos \varphi
_{p}\cos \varphi _{m})\nonumber\\
&+&\alpha E_{\rm J}\left[1-\cos \left(2\pi
f+2\varphi_{m}\right)\right],
\end{eqnarray}
with the reduced magnetic flux $f=\Phi_{e}/\Phi_{0}$ and the
magnetic flux quantum $\Phi_{0}$. The third term $I\Phi(t)$ in
Eq.~(\ref{eq:1}) plays the similar role as the electric-dipole
interactions between the nature atoms and the electric fields, and
describes the interaction between the SFQ circuit and the
time-dependent magnetic flux provided by the external circuit. The
parameter $I$ in Eq.~(\ref{eq:1}) denotes the loop current of the
SFQ circuit given by~\cite{liu2006}
\begin{equation}\label{eq:3}
I=\frac{\alpha\,I_{0}}{2\alpha+1}\left[-2\sin\varphi_{m}\cos\varphi_{p}
+\sin\left(2\pi f+2 \varphi_{m}\right)\right],
\end{equation}
when the time-dependent magnetic flux $\Phi(t)=0$, here $I_{0}=2\pi
E_{\rm J}/\Phi_{0}$.  We note that Eq.~(\ref{eq:3}) is different
from those in the former literatures (e.g., in
Refs.~\cite{liu2005,youjq}). This difference results from different
phase transformations
\begin{eqnarray}
\phi_{p}&=&\frac{1}{2}(\phi_{1}+\phi_{2}), \label{eq:4a}\\
\phi_{m}&=&\frac{1}{2}(\phi_{2}-\phi_{1})+\frac{2\pi\alpha
}{(1+2\alpha)}\frac{\Phi(t)}{\Phi_{0}},\label{eq:4b}
\end{eqnarray}
with the superconducting phase differences $\phi_{1}$ and $\phi_{2}$
of the two identical Josephson junctions. We also use the phase
constraint condition for superconducting phase differences
$\phi_{i}$ (with $i=1,\,2,\,3$) of the three Josephson junctions as
\begin{equation}
-\phi_{1}+\phi_{2}+\phi_{3}+\frac{2\pi\Phi_{e}}{\Phi_{0}}+\frac{2\pi\Phi(t)}{\Phi_{0}}=0,
\end{equation}
when Eqs.~(\ref{eq:2u}) and (\ref{eq:3}) are derived.

In the former literatures (e.g., in Refs.~\cite{liu2005,youjq}), the
second term in Eq.~(\ref{eq:4b}) for the phase transformations has
been neglected. Our derivation here and the former derivation in the literatures~\cite{liu2005,youjq}
can give the same type of the interaction Hamiltonian between the time-dependent magnetic flux
and the superconducting flux qubit circuit,
but there is a significant difference. In the derivation of Refs.~\cite{liu2005,youjq}, $I$ in Eq.~(\ref{eq:1}) is  the supercurrent passing through one of the three Josephson junctions. But for our derivation here, $I$ in Eq.~(\ref{eq:1}) is the loop current of
the SFQ circuit. Thus two derivations result in different coupling strengths between the SFQ circuit and
the external magnetic flux. We think that the phase
transformations in Eqs.~(\ref{eq:4a}) and (\ref{eq:4b}) are more
appropriate than those used in the former literatures (e.g., in
Refs.~\cite{liu2005,youjq}) when the time-dependent magnetic flux is
considered. Because in this transformation, the interaction between
the SFQ circuit and the external circuit can be expressed as the
product~\cite{liu2006} of the loop current $I$ of the SFQ circuit
and the external magnetic flux $\Phi(t)$ provided by the external
circuit. This is in accordance with the interaction energy between the circulating current in a metal loop
and the external magnetic flux.

We know that the loop current of the SFQ circuit equals to the summation of the supercurrent and the displacement current
through one of the three Josephson junctions.
The phase transformations in the former literatures~\cite{liu2005,youjq}
result in an approximated current-flux interaction Hamiltonian between the SFQ circuit and
the external magnetic flux $\Phi(t)$, because the displacement currents in the three Josephson junctions are simply
neglected.
However, the phase transformations, used in Eqs.~(\ref{eq:4a}) and (\ref{eq:4b}),
result in that $I$ in Eq.~(\ref{eq:1}) is just the loop current
given in Eq.~(\ref{eq:3}), which is calculated by including both the
supercurrent and the displacement current for each junction. Thus the phase transformations, used in Eqs.~(\ref{eq:4a}) and (\ref{eq:4b}), overcome the drawback of former derivations~\cite{liu2005,youjq}
and are more reasonable.

We should also note that the transformations applied in
Eqs.~(\ref{eq:4a}) and (\ref{eq:4b}) do not change the basic results
on the selection rules and the adiabatic control of the quantum
states that were studied in Ref.~\cite{liu2005} in which
the displacement currents are neglected. This can be very easily verified by using
Eq.~(\ref{eq:2u}) and Eq.~(\ref{eq:3}), that is: (i) when $f=0.5$
which is called as an optimal point, the potential energy in
Eq.~(\ref{eq:2u}) is an even function of the variables $\phi_{p}$ and
$\phi_{m}$,  and the loop current in Eq.~(\ref{eq:3}) is an odd
function of the variables $\phi_{m}$ and $\phi_{p}$. Therefore the
potential energy and the loop current have the well-defined
symmetry. (ii) When $f \neq 0.5$, the inversion symmetries for both
the potential energy in Eq.~(\ref{eq:2u}) and the loop current
in Eq.~(\ref{eq:3}) do not exist. Therefore, the conclusions in both (i) and (ii)
are the same as those in Refs.~\cite{liu2005,youjq}. However due to the different expressions
of the loop currents in Eq.~(\ref{eq:3}) and in
Refs.~\cite{liu2005,youjq}, the transition matrix elements will be
re-normalized. Below we will further clarify this conclusion via the
discussions on the selection rules and numerical calculations for
the transition matrix elements.

\begingroup \squeezetable
\begin{table}
\caption{\label{tab1} Comparison between SFQ circuits and natural
atoms for dipole moments, parities, symmetry, and selection rules.}
\begin{ruledtabular}
\begin{tabular}{c|c|c|p{1cm}|c}
Atoms& Dipole moments & Parities & Symmetry  & Selection rules  \\
\hline
 Natural atoms & $\propto e \overrightarrow{r}$ & Odd& Well-defined & Have\\
\hline SQC ($f=0.5$) & $\begin{array}{c}\propto -\sin(2\varphi_{m})\\
-2\sin \varphi_{m}\cos\varphi_{p}
\end{array}$& Odd&
Well-defined &Have\\
\hline \ SQC ($f\neq 0.5$)  & $\begin{array}{c}\propto\sin
(2\varphi_{m}+2\pi f)\\
-2\sin \varphi_{m}\cos\varphi_{p}\end{array}$ & No
parity& Broken& Not have\\
\end{tabular}
\end{ruledtabular}
\end{table}
\endgroup

\begin{figure}
\includegraphics[bb=175 254 460 470, width=8 cm, clip]{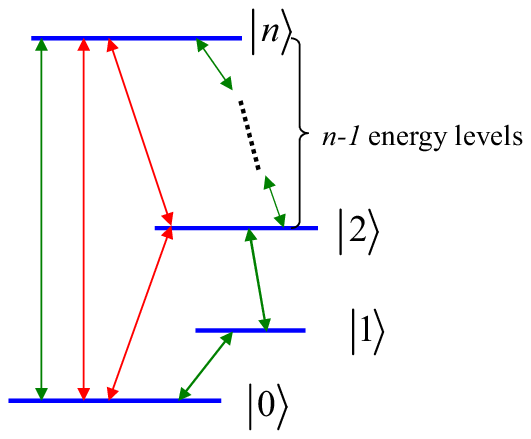}
\caption[]{(Color online) Schematic diagram for the coexistence of
different photon transition processes when the inversion symmetry of
the potential energy is broken, i.e., $f\neq 0.5$. In this case, all
of the transitions between any two energy levels are possible, i.e., there
is no forbidden transition. For example, the loop formed by three
red arrow lines (which link to ground state, the second excited
state, and the $n$th excited state) denotes a coexistence of single-
and two-photon processes for the $(n+1)$-level system. The
coexistence of single- and two-photon processes can also be formed
by the ground, first excited, and second excited states. However,
the loop formed by the green arrow lines (which link the ground,
second, third, until $n$th excited states) denotes a coexistence
of single- and $n$-photon processes for the $(n+1)$-level system.
This schematic diagram also shows that many different photon
processes can coexist in the SFQ circuit with the broken inversion
symmetry.} \label{fig2}
\end{figure}

\subsection{Selection rules and coexistence of multi-photon processes in $n$-level systems}

As a necessary supplementary and generalization of the results in
Ref.~\cite{liu2005} for the microwave-induced transitions between
two different energy levels, we now rewrite the Hamiltonian in
Eq.~(\ref{eq:1}) using eigenstates $\{|i\rangle,\, i=0,\,\cdots n\}$
of the SFQ circuits as the basis
\begin{equation}\label{eq:6}
H=\sum_{i}\hbar\omega_{ii}|i\rangle\langle
i|+\sum_{i,j=0}^{n}I_{ij}|i\rangle\langle j|\Phi(t)\;,
\end{equation}
with ``dipole" matrix elements $I_{ij}=\langle i|I|j\rangle$ and
the eigenvalue $\hbar\omega_{ii}$ of the eigenstate $|i\rangle$.

As discussed above for $f=0.5$, the potential energy in
Eq.~(\ref{eq:2u}) and the loop current in Eq.~(\ref{eq:3}) have
inversion symmetries, and also all eigenstates of the SFQ circuit
have well-defined parities. The loop current  $I$ in
Eq.~(\ref{eq:3}) is an odd function of the variables $\phi_{p}$ and
$\phi_{m}$. Therefore, at the point $f=0.5$, the SFQ circuit has the
same selection rules as the natural atoms, and the microwave-induced
transition can only link two states which have different parities.
However, the symmetry is broken when $f\neq 0.5$. Thus the selection rules
of the SFQ circuits do not exist, and the microwave-induced transitions
between any two energy levels are possible. The comparison of the
selection rules between the SFQ circuits and natural atoms is
summarized in Table~\ref{tab1}.

To compare the $f$-dependent
transition matrix elements using the loop current in
Eq.~(\ref{eq:3}) with those in, e.g., Ref.~\cite{liu2005},  the
moduli of the transition matrix elements $I_{ij}=\langle i|I|j\rangle$ versus the reduced
magnetic flux $f$ is plotted in Fig.~\ref{fig1} for the three lowest
energy levels with $i,\,j=0,\,1,\,2$ and $i\neq j$. Numerical
results in Fig.~\ref{fig1} clearly show that there is the same
transition rule for the current operator used in
Eq.~(\ref{eq:3}) and in Ref.~\cite{liu2005}. However, as shown in
Fig.~\ref{fig1}, the amplitudes of the transition matrix elements
are different for the two different current expressions in
Eq.~(\ref{eq:3}) and in Ref.~\cite{liu2005}.
It is obvious that the coupling strength given by our theoretical derivation here
is bigger than that given in Ref.~\cite{liu2005} near the optimal point.

\begin{figure}
\includegraphics[bb=60 225 580 460, width=8.5 cm,clip]{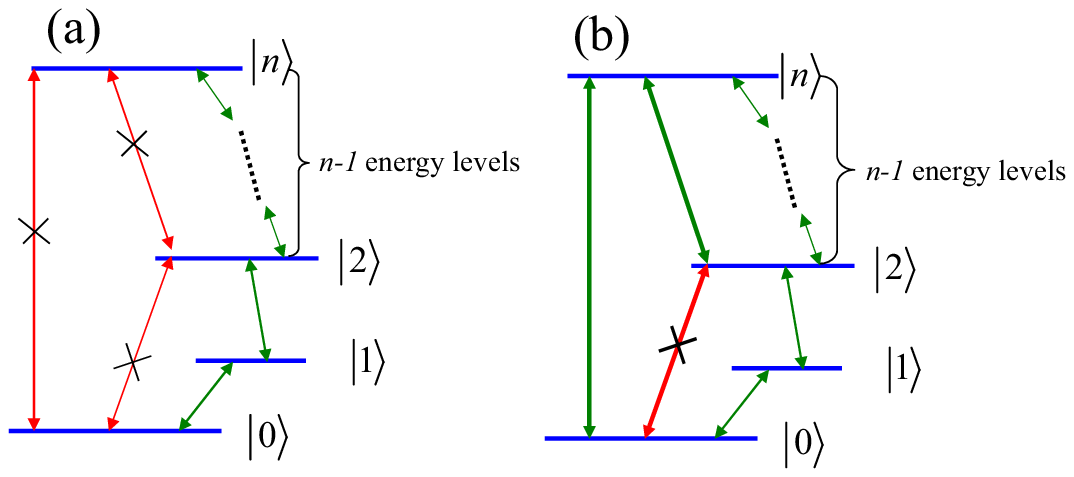}
\caption[]{(Color online) The Schematic diagram for the
photon transition processes when the inversion symmetry of the
potential energy is well-defined, i.e., $f=0.5$. For convenience, let us assume that the microwave-induced transition
between the state $|n\rangle$ and the state $|n+1\rangle$ is possible for
the $(n+1)$-level system. Therefore, it is clear that the transition
between the state $|0\rangle$ and the state $|2\rangle$ is
prohibited. In (a), if the transition between the state $|2\rangle$
and the state $|n\rangle$ is prohibited, then the transition between
the state $|n\rangle$ and the state $|0\rangle$ is also prohibited;
In (b), if the transition between the state $|2\rangle$ and the
state $|n\rangle$ is allowed, then the transition between the state
$|n\rangle$ and the state $|0\rangle$ is also allowed, however
the transition between the state $|0\rangle$ and the state $|2\rangle$ is
forbidden. In both figures, the sign ``$\times$" denotes that the
electric-dipole-like microwave-induced transition is prohibited.
Because the parities for those states are the same. }
\label{n-level}
\end{figure}

As shown in Table~\ref{tab1} and Fig.~\ref{fig1}, when the inversion
symmetry of the potential energy is broken (i.e., $f\neq 0.5$), all
transition matrix elements are nonzero, thus the transitions
between any two levels are allowed. In this case, the single-
and $n$-photon processes can coexist for the $(n+1)$-level SFQ
circuit. That is, for an $(n+1)$-level system, the transition from the
ground state $|0\rangle$ to the $n$th excited state $|n\rangle$ can
be realized by either the single-photon process ($|0\rangle
\rightarrow |n\rangle$) or the $n$-photon processes
($|0\rangle\rightarrow|1\rangle\rightarrow
\cdots|n-1\rangle\rightarrow|n\rangle$). Similarly, many different
photon processes can also coexist in the case with the broken
inversion symmetry. The coexistence of single- and $n$-photon
processes has been schematically shown in Fig.~\ref{fig2}. When
$n=2$, we have the three-level SFQ circuit as discussed in
Ref.~\cite{liu2005}, then the single- and two-photon processes can coexist. It
should be noted that the transitions between two energy levels
should obey the selection rules at the optimal point $f=0.5$. The
photon transition processes for the $(n+1)$-level SFQ circuit have
been schematically shown in Fig.~\ref{n-level} for the case
$f=0.5$.

In above, we mainly analyze the basic properties of the
multi-level SFQ circuits. Below we will focus on new features of the
SFQ (or quantum two-level system formed by the superconducting flux
qubit circuit) when the inversion symmetry of the potential energy
is broken.

\begin{figure}
\includegraphics[bb=20 15 280 210, width=8 cm, clip]{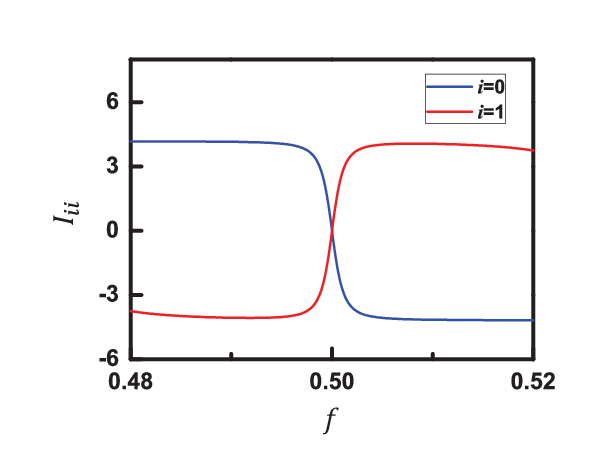}
\caption[]{(Color online)
The $f$-dependent loop current $I_{ii}=\langle
i|I|i\rangle$ for the two lowest energy levels $|0\rangle$ and
$|1\rangle$.  Here, we take the typical numbers $\alpha=0.8$
and $E_{\rm J}=40 E_{\rm c}$ in our numerical simulations.}
\label{fig3}
\end{figure}

\section{New phenomena induced by longitudinal couplings between SFQs and
external magnetic fluxes}

\subsection{Theoretical model on the couplings between SFQs and
external magnetic fluxes}

Let us consider the case of the two lowest energy levels for the SFQ
circuits, i.e., $n=1$. In this case, the Hamiltonian in
Eq.~(\ref{eq:6}) is reduced to that of the superconducting flux
qubits (SFQs), driven by the time-dependent external magnetic flux.
In Fig.~\ref{fig3}, the matrix elements of the loop current $I$ of
the SFQs for the two lowest energy levels $|0\rangle$ and $|1\rangle$ are
plotted versus the reduced magnetic flux $f$. Fig.~\ref{fig3} shows
a well-known result that the loop current $I$ in the ground
$|0\rangle$ and the first excited $|1\rangle$ states are zero at the
optimal point $f=0.5$. However, once the symmetry is broken, the
loop current for both states are nonzero. We note that the loop current  $I_{ii}$
also depends on the ratios $\alpha=E_{\rm  J3}/E_{\rm  J1}$ and $E_{\rm J}/E_{\rm  c}$.  As shown in Fig.~\ref{fig3}, the loop current
$I_{ii}$ is almost a constant for $\alpha=0.8$ and $E_{\rm J}/E_{\rm c}=40$  when $f$ deviates from $0.5$, however it is not always the case
for other parameter ratios.

Based on the above discussions, for a SFQ
interacting with the time-dependent magnetic flux, we have the
following general Hamiltonian
\begin{eqnarray}\label{eq:7}
H&=&\sum_{i=0}^{1}\hbar\omega_{ii}(f)|i\rangle\langle
i|+\left[I_{01}(f)|0\rangle\langle 1|+I_{10}(f)|1\rangle\langle
0|\right]\Phi(t)\nonumber\\
&+&\left[I_{00}(f)|0\rangle\langle 0|+I_{11}(f)|1\rangle\langle
1|\right]\Phi(t).
\end{eqnarray}
Here, we write $\omega_{ii}(f)$ and $I_{ij}(f)$ (with $i,\,j=0,\,1$)
to emphasize the $f$-dependent parameters.

Let us now first discuss the interaction between the SFQ and the
classical magnetic flux $\Phi(t)$ using the Hamiltonian in
Eq.~(\ref{eq:7}) when $f=0.5$. The above analytical analysis
together with Fig.~\ref{fig1} and Fig.~\ref{fig3} show
\begin{equation}\label{eq:8}
I_{00}(f=0.5)=I_{11}(f=0.5)=0,
\end{equation}
and
\begin{equation}\label{eq:9}
I_{10}(f=0.5)=I_{01}(f=0.5)\neq 0.
\end{equation}
Therefore, in Eq.~(\ref{eq:7}), the two coupling terms become
\begin{equation}
I_{ii}(f=0.5)|i\rangle\langle i| \Phi(t)=0,
\end{equation}
with $i=0,\,1$, which means that there is no longitudinal
coupling between the time-dependent magnetic flux and the SFQ at the
optimal point. There are only coupling terms
$I_{01}(f=0.5)(|0\rangle\langle 1|+|1\rangle\langle 0|) \Phi(t)$ in
Eq.~(\ref{eq:7}), called as the transverse coupling between the
time-dependent magnetic flux and the SFQ. Therefore, under the
rotating wave approximation, the Hamiltonian in Eq.~(\ref{eq:7}) at
the optimal point can further be reduced to the Jaynes-Cumming
model, which has been extensively explored in the quantum optics and
the circuit QED system.

When $f\neq 0.5$, all elements $I_{ij}(f)$ with $i=0,\,1$ are nonzero.
In this case, the interaction Hamiltonian between the time-dependent
magnetic flux and the SFQs includes both the transverse and longitudinal
couplings, which are less studied.  This longitudinal coupling can
induce some unusual phenomena which will be explored below. For
convenience, our studies below just consider the
case of the longitudinal and transverse couplings between a
driving classical field and a SFQ, however all of the discussions in
the subsections~\ref{3a},~\ref{3b}, and~\ref{3c} can be applied to
the case with many driving fields.

\subsection{Longitudinal coupling induced coexistence of single- and multi-photon
processes in SFQs}\label{3a}

For convenience, using the relations in
Eqs.~(\ref{eq:8}) and (\ref{eq:9}), the Hamiltonian in
Eq.~(\ref{eq:7}) can be rewritten as
\begin{equation}\label{eq:5}
H=\hbar\frac{\omega_{q}}{2}\sigma_{z}+\hbar(\lambda_{x}\sigma_{x}
+\lambda_{z}\sigma_{z})\cos(\omega_{0} t),
\end{equation}
with $\sigma_{z}=|1\rangle\langle 1|-|0\rangle\langle 0|$ and
$\sigma_{x}=|0\rangle\langle 1|+|1\rangle\langle 0|$. Here, we
assume the magnetic flux $\Phi(t)$ in Eq.~(\ref{eq:7}) to be
$\Phi(t)=\Phi \cos(\omega_{0}t)$, and then $\lambda_{x}=\Phi I_{01}/\hbar$
and $\lambda_{z}=\Phi (I_{11}-I_{00})/(2\hbar)$.

If the SFQ works at the optimal point (i.e., $f=0.5$), then
$I_{11}(f=0.5)=0$ which implies the longitudinal coupling constant
$\lambda_{z}=0$. In this case, Eq.~(\ref{eq:5}) becomes  a
standard Hamiltonian of a driven SFQ, and there is only
single-photon resonant transition in the SFQ induced by the external
magnetic flux with the condition $\omega_{q}=\omega_{0}$.

When the reduced magnetic flux deviates from the optimal point,
i.e., $f\neq 0.5$, there are both the transverse and longitudinal
couplings between the SFQ and the external magnetic flux. In
contrast to the case of only the single-photon process for the
transverse coupling between the SFQ and the external magnetic flux,
the longitudinal coupling can result in the coexistence of the
single- and multi-photon processes in the SFQs. To demonstrate this,
we now apply a unitary transformation
\begin{equation}\label{eq:11b}
U(t)=\exp\left[-\frac{i}{2}\left(\omega_{0}t
+2\frac{\lambda_{z}}{\omega_{0}}\sin\omega_{0}t\right)\sigma_{z}\right]
\end{equation}
to Eq.~(\ref{eq:5}), and thus the Hamiltonian in Eq.~(\ref{eq:5})
becomes
\begin{eqnarray}\label{eq:11a}
H&=&\hbar\frac{\omega_{q}-\omega_{0}}{2}\sigma_{z}+
\hbar\sum_{n}\left[\lambda_{n}e^{-in\omega_{0}t}\sigma_{+}+{\rm
h.c.}\right],
\end{eqnarray}
under the rotating wave approximation. The effective Rabi frequency
in Eq.~(\ref{eq:11a}) is given by
\begin{equation}\label{eq:13}
\lambda_{n}=\lambda_{x} J_{n}\left[\frac{2\lambda_{z}}{\omega_{0}}
\right],
\end{equation}
which depends on both $\lambda_{z}$ and $\omega_{0}$ with the Bessel
functions $J_{n}(2\lambda_{z}/\omega_{0} )$ of the first kind. When
deriving Eq.~(\ref{eq:11a}), we use the relation
\begin{equation}
\exp\left[i\frac{2\lambda_{z}}{\omega_{0}}\sin(\omega_{0}
t)\right]=\sum_{n=-\infty}^{n=\infty}J_{n}\left[\frac{2\lambda_{z}}{\omega_{0}}
\right]\exp\left[i n\omega_{0}t\right].
\end{equation}

\begin{figure}
\includegraphics[bb=180 370 370 510, width=7 cm, clip]{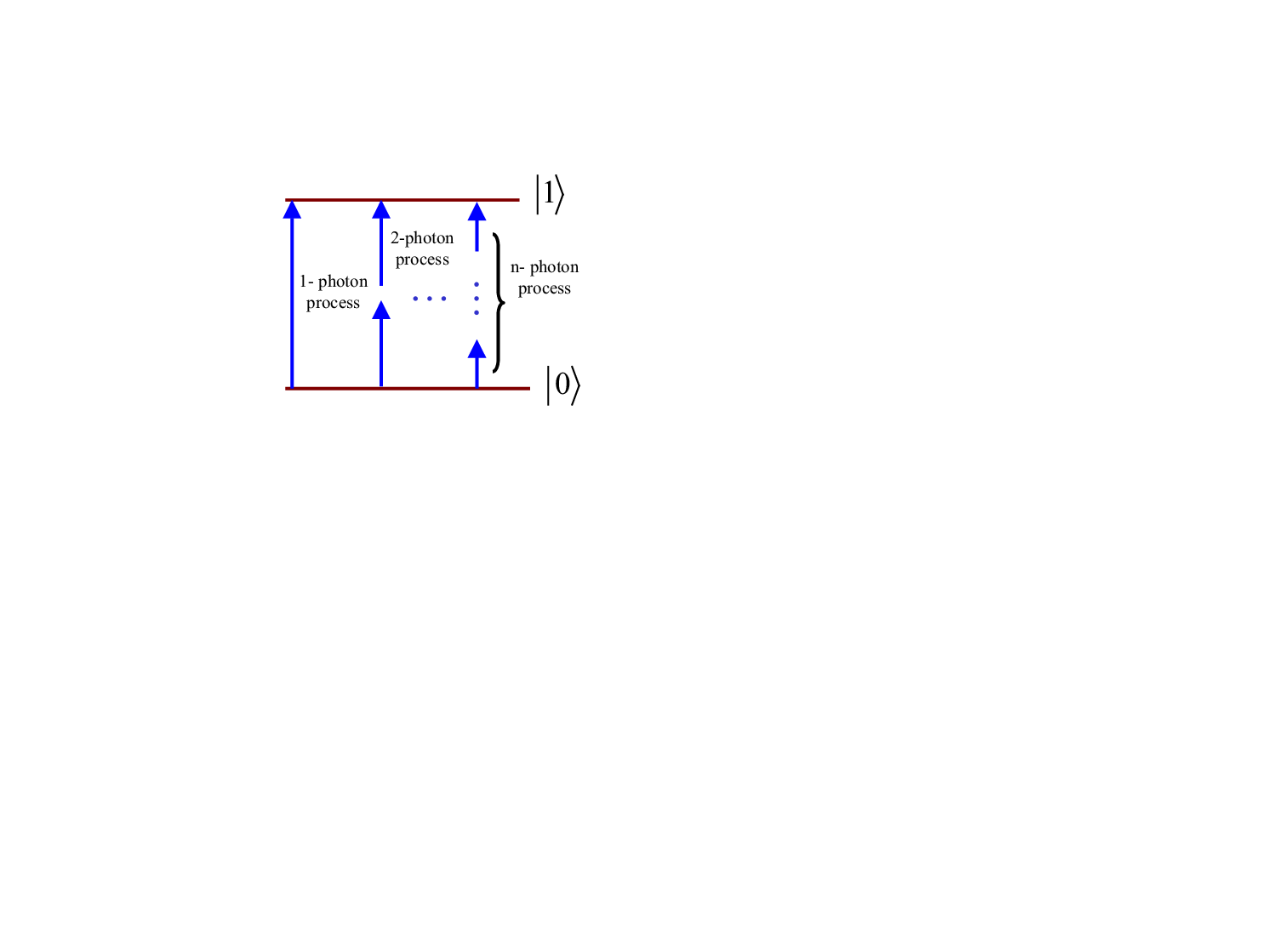}
\caption[]{(Color online) Schematic diagram for the longitudinal
coupling induced coexistence of multi-photon processes in the
SFQs}\label{fig4}
\end{figure}

We note if the SFQ works at the optimal point, then $\lambda_{z}=0$
and the Bessel functions
\begin{eqnarray}
J_{n\neq 0}\left(\frac{2\lambda_{z}}{\omega_{0}}=0\right)&=&0,\\
J_{n=0}\left(\frac{2\lambda_{z}}{\omega_{0}}=0\right)&=&1.
\end{eqnarray}
In this case, Eq.~(\ref{eq:11a}) is reduced to the usual
Jaynes-Cumming model for the externally driven two-level system, and
only describes the single-photon resonant transition with the
condition $\omega_{q}=\omega_{0}$. The single-photon process is
characterized by the term for $n=0$ in Eq.~(\ref{eq:11a}) with the
amplitude
\begin{equation}
\lambda_{0}=\lambda_{x}
J_{0}\left(\frac{2\lambda_{z}}{\omega_{0}}=0\right)=\lambda_{x}.
\end{equation}
This is also an obvious result of Eq.~(\ref{eq:5}) for
$\lambda_{z}=0$ and with the rotating wave approximation as
described above.

However, for the case $\lambda_{z}\neq 0$,  all of the Bessel functions
$J_{n}(2\lambda_{z}/\omega_{0} )$ and then $\lambda_{n}$ are nonzero
except some special ratios $2\lambda_{z}/\omega_{0}$, which are
roots of the Bessel functions. For nonzero $\lambda_{n}$, Eq.~
(\ref{eq:11a}) shows an obvious resonant condition
\begin{equation}
\omega_{q}=(n+1)\omega_{0}.
\end{equation}
Therefore when the inversion symmetry is broken, the longitudinal
coupling can induce the coexistence of single- and multi-photon
processes in the SFQ. The coexistence of  single- and
multi-photon processes in the SFQs is schematically shown in
Fig.~\ref{fig4}.

In summary, the conditions for the coexistence of single- and
multi-photon processes in the SFQs are: (a) the SFQs do not work at
the optimal point, and thus there are both the longitudinal and
transverse couplings between the SFQs and the external magnetic
fluxes; (b) the ratios $2\lambda_{z}/\omega_{0}$ are not roots of
the function $J_{n}(2\lambda_{z}/\omega_{0})$. The multi-photon
processes in the SFQs with the driving fields~\cite{large2} have
been experimentally observed (e.g., in
Refs.~\cite{multi1,multi2,large1,berns,large3}). Thus the above two
conditions and our studies here should be necessarily theoretical
complementary to these experimental studies (e.g., in
Refs.~\cite{multi1,multi2,large1,berns,large3}).

We notice that the Hamiltonian used in Eq.~(\ref{eq:5}) is
equivalent to one, commonly used in the literatures (e.g.,
Refs.~\cite{multi1,multi2}) for SFQs, described by
\begin{equation}\label{eq:10}
H^{\prime}=\varepsilon\sigma_{z}+\Delta
\sigma_{x}+\lambda\sigma_{z}\cos(\omega_{0} t),
\end{equation}
in the loop current basis. Because Eq.~(\ref{eq:10}) can be easily
transformed to Eq.~(\ref{eq:5}) in the qubit basis by
diagonalizing the first two terms of Eq.~(\ref{eq:10}).
We also note that the longitudinal coupling induced coexistence of multi-photon
processes in SFQs actually bears similarity to the phenomenon of coherent
destruction of tunneling~\cite{CDT1,CDT2}.

\subsection{Longitudinal coupling induced transparency of the SFQs to the
transverse coupling fields}\label{3b}

We now show how the longitudinal coupling field can result in the
transparency of the SFQ to the transverse coupling fields. Let us
first give the solutions of the Hamiltonian in Eq.~(\ref{eq:11a}).
We assume that the solutions $|\Psi\rangle$ of Eq.~(\ref{eq:11a})
have the following form
\begin{equation}\label{eq:19}
|\Psi\rangle=A(t)|0\rangle+B(t)|1\rangle.
\end{equation}
According to the expansion of the Hamiltonian in Eq.~(\ref{eq:11a}), there are many
resonant peaks under the frequency matching condition
$\omega_{q}=(n+1)\omega_{0}$. If we assume that the frequency of the
driving field satisfies the condition $\omega_{0}=\omega_{q}/(n+1)$,
then the time-dependent parameters $A(t)$ and $B(t)$ can be given by
\begin{eqnarray}
A(t)&=& \left\{A(0)\left[\cos\left(\frac{\Omega_{n}
t}{2}\right)-i\frac{\Delta_{n}}{\Omega_{n}}\sin\left(\frac{\Omega_{n}
t}{2}\right)
\right] \right.\nonumber\\
&-&\left. i
B(0)\frac{2\lambda_{n}}{\Omega_{n}}\sin\left(\frac{\Omega_{n}
t}{2}\right)\right\}\exp\left[i\frac{n\omega_{0} t}{2}\right],\label{eq:21}\\
B(t)&=&\left\{B(0)\left[\cos\left(\frac{\Omega_{n}
t}{2}\right)+i\frac{\Delta_{n}}{\Omega_{n}}\sin\left(\frac{\Omega_{n}
t}{2}\right)
\right] \right.\nonumber\\
&-&\left. i
A(0)\frac{2\lambda_{n}}{\Omega_{n}}\sin\left(\frac{\Omega_{n}
t}{2}\right)\right\}\exp\left[-i\frac{n\omega_{0}
t}{2}\right].\label{eq:22}
\end{eqnarray}
Here, $A(0)$ and $B(0)$ are given by the initial conditions of
Eq.~(\ref{eq:19}). The Rabi frequency $\Omega_{n}$ and the parameter
$\Delta_{n}$ are given by
\begin{eqnarray}
\Omega_{n}&=&\sqrt{\Delta_{n}^2+4|\lambda_{n}|^2},\\
\Delta_{n}&=&\omega_{q}-(n+1)\omega_{0}.\label{eq:25b}
\end{eqnarray}

If the SFQ is initially prepared to the ground state, i.e.,
$A(0)=1$, and also the coupling constant $\lambda_{n}$ satisfies the
condition $\lambda_{n}=0$,  then from
Eqs.~(\ref{eq:21}) and (\ref{eq:22}), we can obtain  $|B(t)|\equiv
0$ and $|A(t)|\equiv 1$. In this case, the SFQ is always in its
ground state and the population in the excited state is always zero
even that the resonant condition $\omega_{q}=(n+1)\omega_{0}$ is
satisfied. This means that the SFQ is transparent to the transverse field due to
the longitudinal coupling. However, when the
longitudinal coupling is zero, once the resonant condition is
satisfied, the transverse field can be absorbed by the SFQ.

\subsection{Longitudinal coupling induced dynamical quantum Zeno
effect}\label{3c}

We now study another interesting phenomenon, that is, the environmental
effect of the flux qubit can be switched off  by virtue of the
longitudinal coupling field when the inversion symmetry of the SFQ
potential energy is broken. The Hamiltonian of the driven SFQ
interacting with the environment can be written as
\begin{eqnarray}\label{eq:25}
H_{\rm
en}&=&\hbar\frac{\omega_{q}}{2}\sigma_{z}+\hbar(\lambda_{x}\sigma_{x}
+\lambda_{z}\sigma_{z})\cos(\omega_{0} t)\nonumber\\
&+&\sum_{i}\hbar\omega_{i}b^{\dagger}_{i}b_{i}
+\hbar\sum_{i}(g_{i}\sigma_{+}b_{i}+g^{*}_{i}\sigma_{-}b^{\dagger}_{i}),
\end{eqnarray}
with the definition $\sigma_{x}=\sigma_{+}+\sigma_{-}$ for the
ladder operators $\sigma_{\pm}$. Here, as in Eq.~(\ref{eq:5}), the
longitudinal and transverse couplings between the SFQ and the
classical fields are characterized by the parameters $\lambda_{z}$
and $\lambda_{x}$. The environment is presented as a set of harmonic
oscillators, each with the frequency $\omega_{i}$ and the creation (annihilation)
operator $b_{i}^{\dagger}$ ($b_{i}$). The coupling constant between the SFQ
and the $i$th bosonic mode is denoted by $g_{i}$. If a unitary
transformation as in Eq.~(\ref{eq:11b}) is applied to the
Hamiltonian in Eq.~(\ref{eq:25}), then we have an effective
Hamiltonian
\begin{eqnarray}\label{eq:26}
H_{\rm en}^{\rm
eff}&=&\hbar\frac{\omega_{q}-\omega_{0}}{2}\sigma_{z}+
\hbar\sum_{n}\left[\lambda_{n}e^{-in\omega_{0}t}\sigma_{+}+{\rm
h.c.}\right]\\
&+&\sum_{i}\hbar\omega_{i}b^{\dagger}_{i}b_{i}
+\hbar\sum_{n}\sum_{i}(g_{i}^{(n)}\sigma_{+}b+g^{(n)*}_{i}\sigma_{-}b^{\dagger}),\nonumber
\end{eqnarray}
with
\begin{equation}
g_{i}^{(n)}=g_{i}\,J_{n}\left(\frac{2\lambda_{z}}{\omega_{0}}\right)\exp\left[i(n+1)
\omega_{0} t\right].
\end{equation}
Here, $J_{n}(2\lambda_{z}/{\omega_{0}})$ and $\lambda_{n}$ are given
in Eq.~(\ref{eq:13}).

For a modulation of frequency $\omega_{0}$ with the resonant
$\omega_{q}=(n+1)\omega_{0}$ or near resonant condition
$\omega_{q}\approx(n+1)\omega_{0}$, here
$n=[\omega_{q}/\omega_{0}]-1$ denotes the integer nearest to
$(\omega_{q}/\omega_{0})-1$. Then under the rotating wave
approximation with neglecting fast oscillating
terms~\cite{rotating}, we only take the term with the integer number
$n=[\omega_{q}/\omega_{0}]-1$, and then the Hamiltonian in
Eq.~(\ref{eq:26}) becomes
\begin{eqnarray}\label{eq:26a}
H_{\rm en}^{\rm
eff}&=&\hbar\frac{\omega_{q}-\omega_{0}}{2}\sigma_{z}
+\sum_{i}\hbar\omega_{i}b^{\dagger}_{i}b_{i}\\
&+&\hbar\left[\lambda_{[\frac{\omega_{q}}{\omega_{0}}]-1}
\exp\left\{-i\left(\left[\frac{\omega_{q}}{\omega_{0}}\right]-1\right)\omega_{0}t\right\}\sigma_{+}+{\rm
h.c.}\right]\nonumber\\
&+&\hbar\sum_{i}(g_{i}^{\left([\frac{\omega_{q}}{\omega_{0}}]-1\right)}\sigma_{+}b
+{\rm h.c.}).\nonumber
\end{eqnarray}
Equation~(\ref{eq:26a}) clearly shows that the SFQ is decoupled from
its environment when the ratio $2\lambda_{z} /\omega_{0}$ is one of the
zeros of the Bessel function $J_{[\omega_{q} /\omega_{0}]-1
}\left(2\lambda_{z} /\omega_{0}\right)$,  because  the coupling constants
$\lambda_{[\omega_{q}/\omega_{0}]-1}=0$ and
$g_{i}^{\left([\omega_{q}/\omega_{0}]-1\right)}=0$ at
the zeros of the Bessel functions.
Therefore, in contrast to
the longitudinal coupling induced transparency, if the SFQ is initially prepared to the excited
state, i.e., $B(0)=1$, then $|B(t)|\equiv 1$ and $|A(t)|\equiv 0$.
In this case, the SFQ evolves freely and is always in its excited
state, which is equivalent to a dynamical quantum Zeno
effect~\cite{peres,lanzhou}.

\section{Coupling between a driven SFQ and an LC circuit with low frequency}

\subsection{Theoretical model}

\begin{figure}
\includegraphics[bb=222 300 520 470, width=7 cm, clip]{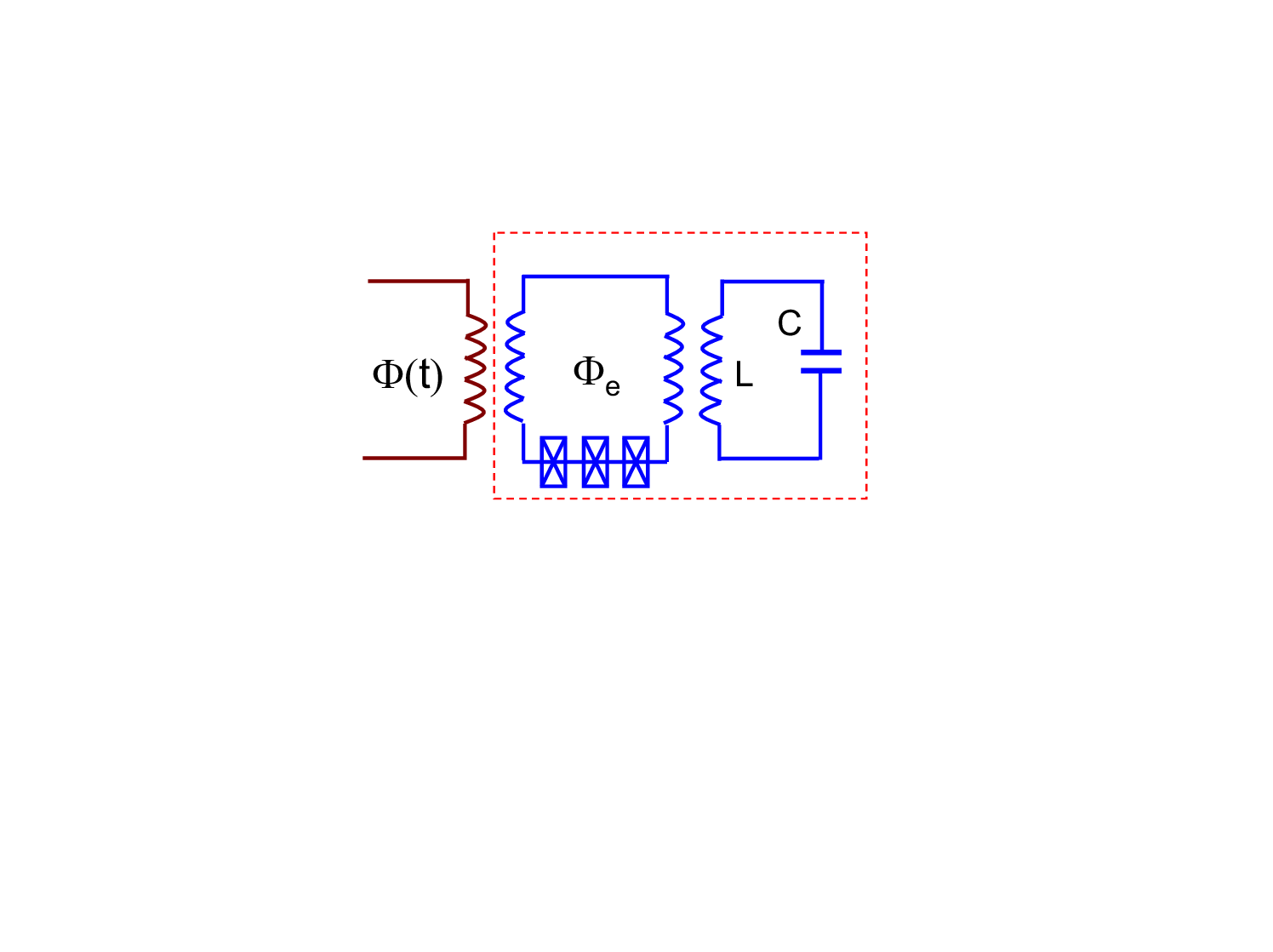}
\caption[]{(Color online) Schematic diagram for the coupling between
an LC circuit and an SFQ (indicated by the red dashed box), which is
driven by the external magnetic flux $\Phi(t)$ (left part of the
figure with the dark red color). }\label{fig6}
\end{figure}

As shown in the red dashed box of Fig.~\ref{fig6}, we first consider
the interaction between a SFQ and a quantized low frequency LC circuit (e.g.,
Refs.~\cite{naturephysics2008,shnirman}) with the frequency $\omega$. The Hamiltonian can be
given by
\begin{equation}\label{eq:28}
H_{q}=\sum_{i=0}^{1}\hbar\omega_{ii}|i\rangle\langle i|+\hbar\omega
a^{\dagger}a+M\sqrt{\frac{\hbar\omega}{2L}}\sum_{i,j=0}^{1}I_{ij}|j\rangle\langle
i|(a+a^{\dagger}),
\end{equation}
which can be further written as
\begin{equation}\label{eq:29}
H_{q}=\hbar\frac{\omega_{q}}{2}\sigma_{z}+\hbar\omega
a^{\dagger}a+\hbar\left(g_{1}\sigma_{x}+g_{2}\sigma_{z}\right)(a+a^{\dagger}),
\end{equation}
with the coefficients
\begin{eqnarray}
g_{1}&=&\frac{M}{\hbar}\sqrt{\frac{\hbar\omega}{2L}}I_{01},\\
g_{2}&=&\frac{M}{2\hbar}\sqrt{\frac{\hbar\omega}{2L}}(I_{11}-I_{00}).
\end{eqnarray}
Here, $M$ is the mutual inductance between the LC circuit and the SFQ.
$L$ is the self-inductance of the LC circuit. $a$ ($a^{\dagger}$) is
the annihilation (creation) operator of the quantized LC circuit. And
the condition $I_{01}=I_{10}$ is used. The transverse and
longitudinal couplings of the SFQ to the LC circuit are realized via
the coupling constants $g_{1}$ and $g_{2}$. The analysis of the
inversion symmetry of the SFQ potential energy tells us that the
longitudinal coupling vanishes (i.e., $g_{2}=0$) only at the optimal
point $f=0.5$, however the transverse coupling is always nonzero.
Below, we will study the case that both transverse and longitudinal
coupling terms are nonzero for the reduced magnetic flux $f\neq
0.5$.

As in experiments~\cite{naturephysics2008,cooling} and also
theoretical studies in Ref.~\cite{shnirman}, we now consider that
the LC circuit and the SFQ are in the regime of the very large
detuning, i.e., the dispersive regime
\begin{equation}
\Delta=\omega_{q}-\omega \gg |g_{1}|.
\end{equation}
In this condition, for the simplicity of the discussions, we take an
approximation
\begin{equation}
 \omega_{q}-\omega \approx \omega_{q}+\omega,
\end{equation}
and apply a unitary transformation $U=\exp(-iS)$ to
Eq.~(\ref{eq:29}) with
\begin{equation}
S=\frac{g_{1}}{\omega_{q}}(a^{\dagger}+a)\sigma_{y},
\end{equation}
then we have an effective Hamiltonian
\begin{eqnarray}\label{eq:33}
H_{q}^{\rm
eff}&=&U^{\dagger}H_{q}U\approx\hbar\frac{\omega_{q}}{2}\sigma_{z}+\hbar\omega
a^{\dagger} a\nonumber\\
&+&\hbar\sigma_{z}\left[g_{2}-
\frac{g^2_{1}}{\omega_{q}}(a^\dagger+a)\right] (a^{\dagger}+a),
\end{eqnarray}
which has been obtained in Ref.~\cite{shnirman}. Here, we only keep
to the first order of $g_{1}/\Delta$. As discussed in
Ref.~\cite{shnirman}, this transform considers that the transverse
coupling affects the SFQ via a second-order longitudinal coupling.

\subsection{Single- and two-photon coupling between the low frequency oscillator and the
driven SFQ}

Let us now discuss the coexistence and disappearance of  single-
and two-photon processes in the driven SFQs when the longitudinal
coupling appears. As shown in Fig.~\ref{fig6}, we assume that a
classical magnetic flux $\Phi(t)$ with the frequency $\omega_{0}$ is
applied to the SFQ, which is coupled to an LC circuit. We consider a
general case for the reduced magnetic flux $f\neq 0.5$. In this
case, the classical field has both the transverse and longitudinal
couplings to the SFQ via the operators $\sigma_{x}$ and
$\sigma_{z}$. The total Hamiltonian can be given by
 \begin{equation}\label{eq:36}
 H_{T}=H_{q}+\hbar(\lambda_{x}\,\sigma_{x}+\lambda_{z}\sigma_{z})\cos(\omega_{0}t).
 \end{equation}
As for obtaining Eq.~(\ref{eq:33}), we first apply the unitary
transformation $U=\exp(-iS)$ to Eq.~(\ref{eq:36}), then we obtain an
effective Hamiltonian
 \begin{equation}\label{eq:37}
 H^{\rm eff}_{T}=H^{\rm eff}_{q}+\hbar(\lambda_{x}\,\sigma_{x}+\lambda_{z}\sigma_{z})\cos(\omega_{0}t).
 \end{equation}
Here, we have neglected the three-body coupling terms between the LC circuit, the SFQ,
and the classical magnetic flux by taking an approximation similar to that
in Ref.~\cite{shnirman}. When Eq.~(\ref{eq:37}) is derived, all of the nonresonant conditions
have to be satisfied. However, in contrast to the Ref.~\cite{shnirman}
where only the transverse coupling term is kept, here we keep
both the longitudinal and transverse couplings between the classical
field and the SFQ. Let us now further apply another unitary
transformation
\begin{equation}
U(t)=\exp\left[\frac{i}{2}\left(\omega_{0}t
+2\frac{\lambda_{z}}{\omega_{0}}\sin\omega_{0}t\right)\sigma_{z}\right]
\end{equation}
to Eq.~(\ref{eq:37}), then we have
\begin{eqnarray}\label{eq:11}
H^{\rm
eff}_{T}&=&\hbar\frac{\omega_{q}-\omega_{0}}{2}\sigma_{z}+\hbar\sigma_{z}\left[g_{2}+
\frac{g^2_{1}}{\Delta}(a^\dagger+a)\right]
(a^{\dagger}+a)\nonumber\\
&+& \hbar\omega a^{\dagger} a
+\hbar\sum_{n}\left[\lambda_{n}e^{-in\omega_{0}t}\sigma_{+}+{\rm
h.c.}\right],
\end{eqnarray}
with the effective Rabi frequency $\lambda_{n}=\lambda_{x}
J_{n}(2\lambda_{z}/\omega_{0})$. For the last term in the righthand side of Eq.~(\ref{eq:11}), only 
the term for $n=0$ with the Rabi frequency $\lambda_{0}=\lambda_{x}
J_{0}(2\lambda_{z}/\omega_{0})$ is time independent.

We assume that $\omega_{q}\approx (n+1)\omega_{0}$. In the rotating reference frame
$V(t)=\exp(-in\omega_{0}\sigma_{z} t/2)$ and using the dressed SFQ
basis, Eq.~(\ref{eq:11}) becomes
\begin{equation}\label{eq:12}
H_{\rm T,R}^{\rm eff}=\hbar\frac{\Omega_{\rm
R}}{2}\sigma_{z}+\hbar\omega a^{\dagger}
a+\hbar(\beta_{1}\sigma_{+}a+\beta_{2}\sigma_{+}a^2+{\rm h.c.}),
\end{equation}
with the dressed qubit frequency
\begin{equation}
\Omega_{\rm
R}=\sqrt{[\omega_{q}-(n+1)\omega_{0}]^2+(2\lambda_{n})^{2}}.
\end{equation}
Here, the fast oscillating terms in Eq.~(\ref{eq:11}) and the anti-rotating terms 
in the dressed SFQ basis have been neglected.
The subscript ``R" indicates the rotating reference frame. The coupling constants $\beta_{1}$ and $\beta_{2}$ are
\begin{eqnarray}
\beta_{1}&=&2\frac{\lambda_{n}}{\Omega_{\rm R}}g_{2},\\
\beta_{2}&=&2\frac{\lambda_{n}g^2_{1}}{\Omega_{\rm R}\Delta}.
\end{eqnarray}
Equation~(\ref{eq:12}) shows that the single-photon process survives
when $\Omega_{\rm R}=\omega$, however two-photon process appears
when $\Omega_{\rm R}=2\omega$.

For the single- and two-photon processes in the driven SFQ shown
above, we should notice: (i) if the driven SFQ works at the optimal
point $f=0.5$, then $g_{2}=0$ and the Hamiltonian in
Eq.~(\ref{eq:29}) is reduced to that of the Jaynes-Cumming model.
In this case, $\beta_{1}=0$ in Eq.~(\ref{eq:12}) and only the two-photon
process exists. However, if $f \neq 0.5$, which corresponds to the
broken inversion symmetry of the SFQ potential energy, then both $g_{1}$
and $g_{2}$ are nonzero. In this case, $\beta_{1}$ and $\beta_{2}$ in
Eq.~(\ref{eq:12}) take nonzero values, thus single- and two-photon
processes can coexist. (ii) Eq.~(\ref{eq:12}) also shows that both
$\beta_{1}$ and $\beta_{2}$ are proportional to the $n$th Bessel
functions $J_{n}(2\lambda_{z}/\omega_{0})$. In the case of zeros of
the $n$th Bessel functions, the transverse coupling via $\sigma_{x}$
between the driven SFQ and the LC circuit is switched off, neither
single-photon process nor two-photon process can be observed in the
driven SFQ. This is an additional condition to obtain the
coexistence of single- and two-photon processes in the driven
SFQ~\cite{naturephysics2008}. Therefore, the ratio between the
longitudinal coupling constant $\lambda_{z}$ and the frequency
$\omega_{0}$ of the driving field determines the coexistence and
disappearance of  single- and two-photon processes. (iii) Due to
the coexistence of single- and two-photon processes when 
the inversion symmetry of the SFQ potential energy is broken, the preparation and
engineering of quantum states of the harmonic oscillator can be
more efficient~\cite{yjzhao}. (iv) As the closing remark of this section, we also
note that the dynamics of the SFQ and the LC oscillator in the
dispersive regime beyond the rotating-wave approximation has been
studied in Ref.~\cite{zueco}. This method can also be applied to the derivation in Eq.~(\ref{eq:33})
when the rotating wave approximation cannot be made.

\section{ Conclusions and Discussions }

In summary, as the necessary complementary and generalization of our
earlier studies~\cite{liu2005}, we first give phase transformations
when the time-dependent microwave is applied, and then study the
microwave-induced transitions between different energy levels in the
multi-level systems formed by the SFQ circuits. We have compared the
selection rules between these  superconducting artificial ``atoms"
and the natural atoms. It is found that the selection rules for such
multi-level systems are the same as those of the multi-level natural
atoms when the reduced bias magnetic flux $f$, applied to the
superconducting loop of the SFQ circuits, is at the optimal point
$f=0.5$. This is because the inversion symmetry of the potential
energy is well defined in this case. However, when the reduced bias
magnetic flux is not at the optimal point, i.e., $f\neq 0.5$, the
superconducting ``atoms" have no selection rules, in this case, the
microwave-induced transitions between any two energy levels are
possible.

The inversion symmetry of the potential energy for the SFQ circuits
is not only important to the multi-level systems, but also important
to the two-level systems (qubits). With the broken inversion
symmetry, there are both the transverse and longitudinal couplings
between the SFQs and the external magnetic fluxes. Compared with
the two-level natural atoms that only have the transverse coupling, the SFQs have
several new phenomena due to the existence of additional longitudinal couplings. For example,
in the two-level natural atoms, the single- and multi-photon
processes cannot coexist due to the well-defined parities of the
eigenstates, however, in the SFQs, the longitudinal coupling can
induce the coexistence of single- and multi-photon processes. We
also demonstrate that the longitudinal coupling can result in: (i)
the transparency of the SFQ to the transverse coupling field; (ii)
the dynamical quantum Zeno effect.

We further study the coupling between the driven SFQ and the LC
circuit with the low frequency. We show that the longitudinal
coupling can result in the coexistence of the single- and two-photon
processes. In contrast, only single-photon processes exist for the case
that the longitudinal coupling is zero. We also obtain the
conditions that the single- and two-photon processes can disappear
even there is longitudinal coupling.

\begingroup \squeezetable
\begin{table}
\caption{\label{tab2} Similarities and differences of superconducting charge, flux, and phase qubits
for the optimal point, selection rules and longitudinal coupling.}
\begin{ruledtabular}
\begin{tabular}{l|l|l|l}
Qubit Type& Optimal point  & Selection rules & Longitudinal coupling\\
\hline
Charge & Have & Have (at optimal point) & Have (not at optimal point)\\
\hline
Flux& Have & Have (at optimal point)  & Have (not at optimal point)\\
\hline
Phase& Not have & Not have   & Have (always)
\end{tabular}
\end{ruledtabular}
\end{table}
\endgroup

As summarized in the table~\ref{tab2}, the phase qubits do not have the selection rules and the optimal point, thus
there is always longitudinal coupling between the phase qubit and the microwave fields. However the charge qubits
have the selection rules and the optimal point, thus there is the longitudinal coupling between the charge
qubits and the microwave fields when the charge qubits do not work at the optimal point.
Therefore, we remark that: (i) all results studied in this paper for the SFQs can be
directly generalized to the superconducting phase qubits~\cite{large-phase,martinis}, these results can also be generalized to the charge~\cite{charge,large-charge1,large-charge2,delsing} qubits when the charge qubits do not work at the optimal point. Because all of these qubit systems can have the Hamiltonians similar to those in Eq.~(\ref{eq:5}) and Eq.~(\ref{eq:29})
under proper conditions. (ii) The transverse and longitudinal couplings between the
charge qubit and the LC circuit in the resonant or
near-resonant case have been studied, e.g., in
Ref.~\cite{large-charge1}. However, some new aspects induced by the
longitudinal coupling are still needed to be further explored in the
near future, e.g., photon state engineering.

We now discuss the experimental feasibilities of our proposal. The observation of the coexistence of multi-photon processes in multi-level systems of the superconducting quantum circuits might exceed current experiments for the high frequency cutoff of the cryogenic amplifier. However, as shown in this paper, the nature of the coexistence of multi-photon processes in  multi-level SFQ circuits and the longitudinal coupling induced coexistence of single- and multi-photon processes in two-level SFQ circuits are the same. Thus, we suggest experimentally observing the latter. This should be experimentally realizable with current technology. We note that the longitudinal coupling induced transparency of the SFQs to the transverse coupling fields can be observed by coupling a SFQ to the open one-dimensional space of a transmission line as for experimental observation of
the Autler-Townes effect~\cite{Abdumalikov} in three-level superconducting quantum circuits. The dynamical quantum Zeno effect can be demonstrated by directly measuring the coherent time of the SFQ when the SFQ deviates from the optimal point and a proper longitudinal field is applied to the SFQ. In current experiments, the coupling strength between the SFQ and the microwave field is just determined by the experimental data. To our knowledge, there is a lack of theoretical fittings of experimental data for the coupling strength between
the SFQ and microwave field. Thus, although we think that
the interaction Hamiltonian between the SFQ and the microwave field in Eq.~(\ref{eq:1}) is more reasonable, the experimental confirmation
is still desirable. This would be very helpful for both theoretical and experimental studies of scalable SFQs in the superconducting quantum information processing.

\section{Acknowledgement}
Y.~X.~Liu is supported by the National Natural Science Foundation of
China under Nos. 10975080, 61025022, and 60836001. X. B. Wang is
supported by the National Basic Research Program of China grant nos
2007CB907900 and 2007CB807901, NSFC grant number 60725416, and China
Hi-Tech program grant no. 2006AA01Z420.


\begin{thebibliography}{99}

\bibitem{p1}Y. Makhlin, G. Sch\"on, and A. Shnirman, Rev. Mod. Phys.~\textbf{73},
357 (2001).

\bibitem{r1}J. Q. You and F. Nori, Phys. Today~\textbf{58} (11), 42 (2005).

\bibitem{r2}R. J. Schoelkopf and S. M. Girvin, Nature~\textbf{451},
664 (2008).

\bibitem{r3}G. Wendin and V. S. Shumeiko, in \emph{Handbook of Theoretical and
Computational Nanotechnology}, edited by M. Rieth and W. Schommers
(American Scientific, California, 2006), Vol.~\textbf{3}.

\bibitem{r4}J. Clarke and F. K. Wilhelm, Nature~\textbf{453}, 1031
(2008).

\bibitem{r5}J. Q. You and F. Nori, Nature~\textbf{474}, 589 (2011).


\bibitem{wallraff}A. Wallraff, D. I. Schuster, A. Blais, L. Frunzio, R. S. Huang,
J. Majer, S. Kumar, S. M. Girvin, and R. J. Schoelkopf, Nature
\textbf{431}, 162 (2004).

\bibitem{mooij1}I. Chiorescu, P. Bertet, K. Semba, Y. Nakamura, C. J. P. M. Harmans,
and J. E. Mooij, Nature~\textbf{431}, 159 (2004).

\bibitem{review1}R. J. Schoelkopf and S. M. Girvin,  Nature~\textbf{451}, 664 (2008).

\bibitem{liudressed}Yu-xi Liu, C. P. Sun, and F. Nori, Phys. Rev. A~\textbf{74},
052321 (2006).

\bibitem{large-charge3} C. M. Wilson, T. Duty, F. Persson, M. Sandberg, G. Johansson,
and P. Delsing, Phys. Rev. Lett.~\textbf{98}, 257003 (2007).

\bibitem{Wilson-PRB}  C. M. Wilson, G. Johansson, T. Duty, F. Persson, M. Sandberg,
and P. Delsing, Phys. Rev. B~\textbf{81}, 024520 (2010).


\bibitem{wallraff2009}J. M. Fink, R. Bianchetti, M. Baur, M. Goppl, L. Steffen,
S. Filipp, P. J. Leek, A. Blais, and A. Wallraff, Phys. Rev.
Lett.~\textbf{103}, 083601 (2009).

\bibitem{greenberg}Ya. S. Greenberg, Phys. Rev. B~\textbf{76}, 104520
(2007).

\bibitem{jena}G. Oelsner, P. Macha, O. V. Astafiev, E. Il¡¯ichev, M. Grajcar, U. H\"ubner,
B. I. Ivanov, P. Neilinger, and H.-G. Meyer, Phys. Rev.
Lett.~\textbf{110}, 053602 (2013).


\bibitem{yang1}C.-P. Yang, S.-I. Chu, and S. Han, Phys. Rev. A~\textbf{67}, 042311 (2003).

\bibitem{yang2}C.-P. Yang, S.-I. Chu, and S. Han, Phys. Rev. Lett.~\textbf{92},
117902 (2004).


\bibitem{orlando2004}K. V. R. M. Murali, Z. Dutton, W. D. Oliver, D. S. Crankshaw,
and T. P. Orlando, Phys. Rev. Lett.~\textbf{93}, 087003 (2004).

\bibitem{orlando2006} Z. Dutton, K. V. R. M. Murali, W. D. Oliver, and T. P. Orlando,
Phys. Rev. B~\textbf{73}, 104516 (2006).

\bibitem{nist}M. A. Sillanp\"a\"a, J. Li, K. Cicak, F. Altomare, J. I. Park,
R. W. Simmonds, G. S. Paraoanu, and P. J. Hakonen, Phys. Rev.
Lett.~\textbf{103}, 193601 (2009).

\bibitem{goan}X. Z. Yuan, H. S. Goan, C. H. Lin, K. D. Zhu,
and Y. W. Jiang, New J. Phys.~\textbf{10}, 095016 (2008).

\bibitem{falci}J. Siewert, T. Brandes, and G. Falci, Phys. Rev. B~\textbf{79},
024504 (2009).

\bibitem{falci-1}J. Siewert, T. Brandes, and G. Falci,
Opt. Commun.~\textbf{264}, 435 (2006).

\bibitem{ian}H. Ian, Yu-xi Liu, and F. Nori, Phys. Rev. A~\textbf{81}, 063823
(2010).

\bibitem{Abdumalikov} A. A. Abdumalikov, Jr., O. V. Astafiev, A. M. Zagoskin, Yu. A.
Pashkin, Y. Nakamura, and J. S. Tsai, Phys. Rev. Lett.~\textbf{104},
193601 (2010).


\bibitem{blais}J. Joo, J. Bourassa, A. Blais, and B. C. Sanders,
Phys. Rev. Lett.~\textbf{105}, 073601 (2010).


\bibitem{wei}W. Z. Jia and L. F. Wei, Phys. Rev. A~\textbf{82}, 013808
(2010).


\bibitem{youjqprl} J. Q. You, Yu-xi Liu, and F. Nori, Phys. Rev. Lett.~\textbf{100},
047001 (2008).

\bibitem{orlandon}S. O. Valenzuela, W. D. Oliver, D. M. Berns, K. K. Berggren,
L. S. Levitov, and T. P. Orlando, Science~\textbf{314}, 1589 (2006).

\bibitem{cooling}M. Grajcar, S. H. W. van der Ploeg, A. Izmalkov, E. Il'ichev,
H.-G. Meyer, A. Fedorov, A. Shnirman, and G. Schon, Nature
Phys.~\textbf{4}, 612 (2008).

\bibitem{side1}Yu-xi Liu, L. F. Wei, J. R. Johansson, J. S. Tsai, and
F. Nori, Phys. Rev. B ~\textbf{76}, 144518 (2007); cond-mat/0509236.

\bibitem{side2}A. Wallraff, D. I. Schuster, A. Blais, J. M. Gambetta, J. Schreier,
L. Frunzio, M. H. Devoret, S. M. Girvin, and R. J. Schoelkopf, Phys.
Rev. Lett.~\textbf{99}, 050501 (2007).

\bibitem{side3}P. J. Leek, S. Filipp, P. Maurer, M. Baur, R. Bianchetti, J. M. Fink,
M. G\"oppl, L. Steffen, and A. Wallraff, Phys. Rev. B~\textbf{79},
180511(R) (2009).


\bibitem{liu2005}Yu-xi Liu, J. Q. You, L. F. Wei, C. P. Sun, and F. Nori,
Phys. Rev. Lett.~\textbf{95}, 087001 (2005).

\bibitem{scullybook}M. O. Scully and M. S. Zubairy,
{\it Quantum Optics} (Cambridge University Press, Cambridge,
England, 1997).


\bibitem{naturephysics2008} F. Deppe, M. Mariantoni, E. P. Menzel, A. Marx,
S. Saito, K. Kakuyanagi, H. Tanaka, T. Meno, K. Semba, H.
Takayanagi, E. Solano, and R. Gross, Nature Phys.~\textbf{4}, 686
(2008).

\bibitem{nano} T. Niemczyk, F. Deppe, M. Mariantoni, E. P. Menzel, E. Hoffmann,
G. Wild, L. Eggenstein, A. Marx, and R. Gross, Supercond. Sci.
Technol.~\textbf{22},  034009 (2009).

\bibitem{youjqprb} J. Q. You, Yu-xi Liu, C. P. Sun, and F. Nori,
Phys. Rev. B~\textbf{75}, 104516 (2007).

\bibitem{nec}O. Astafiev, K. Inomata, A. O. Niskanen, T. Yamamoto, Yu. A. Pashkin,
Y. Nakamura, and J. S. Tsai, Nature~\textbf{449}, 588 (2007).


\bibitem{liu2006} Yu-xi Liu, L. F. Wei, J. S. Tsai, and F. Nori,
Phys. Rev. Lett.~\textbf{96}, 067003 (2006).


\bibitem{plourde}B. L. T. Plourde, J. Zhang, K. B. Whaley, F. K. Wilhelm,
T. L. Robertson, T. Hime, S. Linzen, P. A. Reichardt, C.-E. Wu, and
J. Clarke, Phys. Rev. B~\textbf{70}, 140501 (2004).

\bibitem{plourde-1} B. L. T. Plourde, T. L. Robertson, P. A. Reichardt, T. Hime, S.
Linzen, C.-E. Wu, and J. Clarke, Phys. Rev. B~\textbf{72}, 060506(R)
(2005).

\bibitem{miro1} M. Grajcar, A. Izmalkov, S. H. W. van der Ploeg, S. Linzen,
E. Il'ichev, Th. Wagner, U. Hubner, H.-G. Meyer, A. M. van den
Brink, S. Uchaikin, A. M. Zagoskin, Phys. Rev. B~\textbf{72},
020503(R) (2005).


\bibitem{mooij}P. Bertet, C. J. P. M. Harmans, and J. E. Mooij,
Phys. Rev. B~\textbf{73}, 064512 (2006).

\bibitem{tsai}A. O. Niskanen, Y. Nakamura, and J. S. Tsai,
Phys. Rev. B~\textbf{73}, 094506 (2006).

\bibitem{miro} M. Grajcar, Yu-xi Liu, F. Nori, and A. M. Zagoskin,
Phys. Rev. B~\textbf{74}, 172505 (2006).

\bibitem{franco}S. Ashhab, A. O. Niskanen, K. Harrabi, Y. Nakamura,
T. Picot, P. C. de Groot, C. J. P. M. Harmans, J. E. Mooij, and F.
Nori, Phys. Rev. B~\textbf{77}, 014510 (2008).

\bibitem{tsaiscience}A. O. Niskanen, K. Harrabi, F. Yoshihara, Y. Nakamura,
S. Lloyd, and J. S. Tsai, Science~\textbf{316}, 723 (2007).

\bibitem{harrabi}K. Harrabi, F. Yoshihara, A. O. Niskanen, Y. Nakamura, and J. S. Tsai
Phys. Rev. B~\textbf{79}, 020507(R) (2009).

\bibitem{shnirman}J. Hauss, A. Fedorov, C. Hutter, A. Shnirman, and
G. Sch\"on, Phys. Rev. Lett.~\textbf{100}, 037003 (2008).

\bibitem{shnirman-1} J. Hauss, A. Fedorov, S. Andre, V. Brosco, C. Hutter, R.
Kothari, S. Yeshwanth, A. Shnirman, and G. Sch\"on, New J.
Phys.~\textbf{10}, 095018 (2008).


\bibitem{orlando}T. P. Orlando, J. E. Mooij, Lin Tian, C. H. van der Wal,
L. S. Levitov, S. Lloyd, and J. J. Mazo, Phys. Rev. B~\textbf{60},
15398 (1999).


\bibitem{youjq}J. Q. You, Y. Nakamura, and F. Nori, Phys. Rev. B~\textbf{71},
024532 (2005).



\bibitem{large2}S. Ashhab, J. R. Johansson, A. M. Zagoskin, and F.
Nori, Phys. Rev. A~\textbf{75}, 063414 (2007).


\bibitem{multi1}
S. Saito, M. Thorwart, H. Tanaka, M. Ueda, H. Nakano, K. Semba, and
H. Takayanagi, Phys. Rev. Lett.~\textbf{93}, 037001 (2004).


\bibitem{multi2} A. Izmalkov, M. Grajcar, E.
Il'ichev, N. Oukhanski, T. Wagner, H.-G. Meyer, W. Krech, M. H. S.
Amin, A. Maassen van den Brink, and A. M. Zagoskin, Europhys.
Lett.~\textbf{65}, 844 (2004).

\bibitem{large1}W. D. Oliver, Y. Yu, J. C. Lee, K. K. Berggren, L. S. Levitov,
and T. P. Orlando, Science~\textbf{310}, 1653 (2005).


\bibitem{berns} D. M. Berns, W. D. Oliver, S. O. Valenzuela, A. V. Shytov, K. K.
Berggren, L. S. Levitov, and T. P. Orlando, Phys. Rev.
Lett.~\textbf{97}, 150502 (2006).

\bibitem{large3}X. Wen and Y. Yu, Phys. Rev. B~\textbf{79}, 094529
(2009).

\bibitem{CDT1}F. Grossmann, T. Dittrich, P. Jung, and P. Hanggi,
Phys. Rev. Lett.~\textbf{67}, 516 (1991).

\bibitem{CDT2}G. Della Valle, M. Ornigotti, E. Cianci, V. Foglietti, P. Laporta, and S. Longhi,
Phys. Rev. Lett.~\textbf{98}, 263601 (2007).


\bibitem{rotating}J. Hausinger and M. Grifoni, Phys. Rev. A~\textbf{81}, 022117
(2010).


\bibitem{peres}A. Peres, Am. J. Phys.~\textbf{48}, 931 (1980).

\bibitem{lanzhou}L. Zhou, S. Yang, Yu-xi Liu, C. P. Sun, F. Nori,
Phys. Rev. A~\textbf{80}, 062109 (2009).

\bibitem{yjzhao} Y. J. Zhao~\textit{et al.}, in preparation.


\bibitem{zueco}D. Zueco, G. M. Reuther, S. Kohler, and P. Hanggi,
Phys. Rev. A~\textbf{80}, 033846 (2009).


\bibitem{large-phase} A. Wallraff, T. Duty, A. Lukashenko, and A. V. Ustinov,
Phys. Rev. Lett.~\textbf{90}, 037003 (2003).


\bibitem{martinis}J. M. Martinis, S. Nam, J. Aumentado, and C. Urbina,
Phys. Rev. Lett.~\textbf{89}, 117901 (2002).


\bibitem{charge}Y.  Nakamura, Yu. A. Pashkin, and J. S. Tsai, Nature~\textbf{398},
786 (1999).

\bibitem{large-charge1}Y. Nakamura, Yu. A. Pashkin, and J. S. Tsai, Phys. Rev. Lett.
~\textbf{87}, 246601 (2001).

\bibitem{large-charge2}M. Sillanp\"a\"a, T. Lehtinen, A. Paila, Y. Makhlin,
and P. Hakonen, Phys. Rev. Lett.~\textbf{96}, 187002 (2006).


\bibitem{delsing}A. Aassime, G. Johansson, G. Wendin, R. J. Schoelkopf,
and P. Delsing, Phys. Rev. Lett.~\textbf{86}, 3376 (2001).

\end{thebibliography}
\end{document}